\documentclass[english,aps,prl,twocolumn,showpacs,superscriptaddress,footinbib]{revtex4-1}  % for review and submission
\usepackage{graphicx}  % needed for figures
\usepackage{dcolumn}   % needed for some tables
\usepackage{bm}        % for math
\usepackage{amssymb}   % for math
\usepackage{babel}
\usepackage{verbatim}
\usepackage{amsmath}
\usepackage{setspace}
\usepackage{epstopdf}

\graphicspath{{figures/}}

\begin{document}

\title{Low field nuclear polarization using Nitrogen Vacancy centers in diamonds}

\author{Y. Hovav}
\affiliation{Dept. of Applied Physics, Rachel and Selim School of Engineering, Hebrew University, Jerusalem 9190401, Israel}

\author{B. Naydenov}
\affiliation{Institute of Quantum Optics,Ulm University, Albert Einstein Allee 11, D-89081 Ulm, Germany}
\affiliation{Centre for Integrated Quantum Science and Technology (IQST), Albert Einstein Allee 11, D-89081 Ulm, Germany}

\author{F. Jelezko}
\affiliation{Institute of Quantum Optics,Ulm University, Albert Einstein Allee 11, D-89081 Ulm, Germany}
\affiliation{Centre for Integrated Quantum Science and Technology (IQST), Albert Einstein Allee 11, D-89081 Ulm, Germany}

\author{N.Bar-Gill}
\affiliation{Dept. of Applied Physics, Rachel and Selim School of Engineering, Hebrew University, Jerusalem 9190401, Israel}
\affiliation{Racah Institute of Physics, The Hebrew University of Jerusalem, Jerusalem 9190401, Israel}
\affiliation{Quantum Information Science Program, Canadian Institute for Advanced Research, 661 University Ave., Suite 505, Toronto, Ontario M5G 1M1, Canada}

\date{\today}
\begin{abstract}
It was recently demonstrated that bulk nuclear polarization can be obtained using Nitrogen Vacancy (NV) color centers in diamonds, even at ambient conditions. This is based on the optical polarization of the NV electron spin, and using several polarization transfer methods. One such method is the NOVEL sequence, where a spin-locked sequence is applied on the NV spin, with a microwave power equal to the nuclear precession frequency. This was performed at relatively high fields, to allow for both polarization transfer and noise decoupling. As a result, this scheme requires accurate magnetic field alignment in order preserve the NV properties. Such a requirement may be undesired or impractical in many practical scenareios. 
Here we present a new sequence, termed the refocused NOVEL, which can be used for polarization transfer (and detection) even at low fields. Numerical simulations are performed, taking into account both the spin Hamiltonian and spin-decoherence, and we show that, under realistic parameters, it can outperform  the NOVEL sequence.
\end{abstract}
\maketitle
%%%%%%%%%%%%%%%%%%%%%%%%%%%%%%%%%%%%%%%%%%%%%%%%%%%%%%
%\section{INTRODUCTION}
%%%%%%%%%%%%%%%%%%%%%%%%%%%%%%%%%%%%%%%%%%%%%%%%%%%%%%
Dynamic Nuclear Polarization (DNP) \cite{Abragam1978} has gained renewed interest in recent years \cite{Becerra1993,Ardenkjaer-Larsen2003} due to its ability to dramatically increase the signals in nuclear magnetic resonance spectroscopy (NMR) and imaging (MRI) experiments. This relies on polarization transfer from electrons to their neighboring nuclei via microwave (MW) irradiation.
The DNP process is typically performed at cryogenic temperatures, in order to gain high initial electron Boltzmann polarization. An alternative approach is to use high non-equilibrium polarization \cite{Fischer1969, Kaptein1978}, allowing high polarizations to be achieved even at ambient conditions.
It was recently demonstrated that Nitrogen-Vacancy (NV) color centers in diamond \cite{Gruber1997, Jelezko2006, Schirhagl2014} can be used as such a polarization source, based on the ability to polarize the NV electronic $S=1$ spin to its $|0\rangle$ ground state using optical illumination with 532 nm green light. Such an NV based polarization source offers many advantages: it allows fast electron polarization; it is extremely stable; it can be used in combination with coherent polarization schemes due to the NV spin long coherence times \cite{Bar-Gill2013}; and it relies on a relatively simple experimental setup. Nevertheless, polarization transfer to nuclei outside of the diamond itself still poses many challenges. It was also shown that nuclear polarization can improve NV based measurements, by increasing the NV spin free evolution/Ramsey coherence time \cite{london2013}.

NV based polarization of $^{13}$C nuclei in the diamond can be achieved using the NOVEL (nuclear orientation via electron spin locking) scheme \cite{Brunner1987,Henstra2008}, allowing for nuclear sensing \cite{london2013} and bulk polarization \cite{Scheuer2016}. During this sequence a spin-lock \cite{Slichter1990} (SL) is applied on the NV spin with a SL Rabi frequency equal to the nuclear Larmor frequency, resulting in the rotating-frame/lab-frame Hartmann-Hahn polarization transfer condition. These NV based experiments were performed at relatively high fields (around 5000 Gauss) to allow for efficient SL noise decoupling together with the Hartmann-Hahn condition. This relatively high magnetic field requires precise alignment of the field direction to that of the NV axis, in order to allow for the laser induced NV spin polarization to take place. NV based polarization transfer was also obtained using an aligned magnetic field of 514 Gauss  \cite{Fischer2013}, for which NV-nuclear state mixing occurs in the excited state, or at arbitrarily aligned fields, but based on the interaction of the NV to its close neighboring $^{13}$C nuclei \cite{Alvarez2015}.

In this article we present a method for coherent  polarization transfer to remote nuclei (and therefore possibly also to nuclei outside of the diamond) at low fields ($\ll$ 500 Gauss), for which field alignment is less demanding. This can be of importance for nuclear polarization applications in which field alignment is undesired or impractical, possibly including the use of nano-diamonds. 

\begin{figure}[!t]
	\centering   
	\includegraphics[width = 0.3\textwidth]{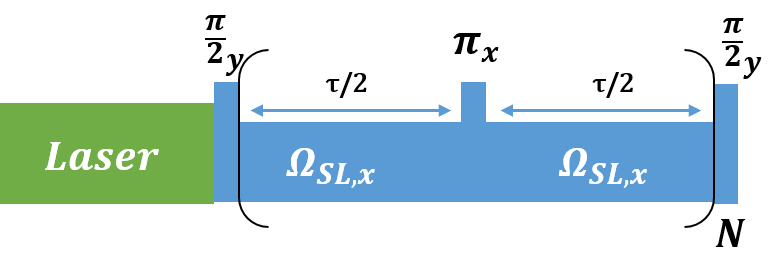}
	\caption{Schematic representation of the refocused-NOVEL sequence, as described in the main text}
	\label{sequence}
\end{figure}

The sequence presented here, is shown schematically in Fig. \ref{sequence}. It is composed of three parts: (1) initialization, which includes a laser pulse for NV polarization and spin state detection, and a $\frac{\pi}{2}$ preparation pulse; (2) the refocused-NOVEL (rNOVEL) sequence, which is composed of $N$ repetitions of two SL pulses of duration $\frac{\tau}{2}$, separated by a $\pi$ pulse of the same phase (unlike the refocused continuous-wave decoupling sequence \cite{Vinther2012,Vinther2013}); and (3), a detection $\frac{\pi}{2}$ pulse, which can be added with different phases for optical detection of the NV polarization. 

In order to describe the effect of this sequence we can consider a simple model spin system composed of the $S = 1$ NV electron, limited to its ${|0\rangle,|-1\rangle}$ subspace, and a spin $I = \frac{1}{2}$ nucleus. The Hamiltonian of this system in the electronic rotating frame is given, assuming ideal $\pi$ pulses for simplicity, by \cite{Slichter1990}:
\begin{align}\label{Spin Hamiltonian}
\begin{split}
H = & \Delta_e S_z - \omega_{n}^0 I_z +  A_\parallel S_z I_z + \frac{1}{2}S_z(A_\perp I^+ +A_\perp^* I^-) +\\
& b(t) S_z + \frac{1}{\sqrt{2}}\Omega_{SL}S_x + \sum_{k = 1,3,5...}^{N}\delta(t-\frac{1}{2}k\tau)\pi_x.
\end{split}
\end{align}
The above Hamiltonian includes the off resonance irradiation on the electronic spin, $\Delta_e$; the nuclear Larmor precession frequency $\omega_n^0$; the secular and pseudo-secular terms of the hyperfine interaction, $A_\parallel$ and $A_\perp$; external pure dephasing noise, b(t); and the MW irradiation term, given by the SL term with an amplitude of $\Omega_{SL}$ and the ideal $\pi_x$ pulses. 

We can next transfer to the interaction frame of the ideal $\pi$ pulses. This is given, for an even $N$, by a unitary transformation of the form  $U(t) =  e^{\int_{0}^{t}i\frac{\pi}{\sqrt{2}}\sum_k\delta(t'-k\tau)S_x dt}$. %?????????????????
The resulting Hamiltonian is then given by: 
\begin{eqnarray}\label{interaction frame Hamiltonian}
%\begin{split}
H^I &=& s(t)[\Delta_e + A_\parallel I_z + \frac{1}{2}(A_\perp I^+ +A_\perp^* I^-) +b(t)] S_z \nonumber \\
&-& \omega_{n}^0 I_z + \Omega_{SL} S_x, \\
&& s(t)= {\left.\begin{array}{rc} \nonumber
	1, & 0 \leq \mod(t,2\tau) < \frac{1}{2}\tau  \\
	-1, & \frac{1}{2}\tau \leq \mod(t,2\tau) < \frac{3}{2}\tau \\
	1, & \frac{3}{2}\tau \leq \mod(t,2\tau) < 2\tau \\
	\end{array} \right\rbrace} .
%\end{split}
\end{eqnarray}
$s(t)$ is a periodic square wave, with a frequency of $\omega_f = \pi/\tau$ (as in CPMG \cite{Meiboom1958} and similar sequences)
which can be expended in terms of a Fourier Series:   
\begin{align}\label{Fourier s(t)}
\begin{split}
 &s(t) = \sum_{k} s_ke^{i k \omega_f t}, \\
 & s_k = 2\frac{(-1)^\frac{k-1}{2}}{\pi k},
 \end{split}
 \end{align}
with $k = \pm1, \pm3,...$. 

In order to simplify the form of $H^I$, we can transfer to a rotating frame in $S_x$, with a frequency of  $k'\omega_f$. Assuming that $\omega_f \gg |\Delta_e+ \frac{1}{2}A_\parallel + b(t)|, |A_\perp|$ (at all times), we can neglect all the $k \neq k'$ time-dependent terms. As such, the Hamiltonian can be written as: 
\begin{align}\label{omega_f rotating frame Hamiltonian}
\begin{split}
&H^R \approx s_k'[\Delta' S_z + \frac{1}{2}(A_\perp I^+ +A_\perp^* I^-)S_z] - 
\omega_{n}^0 I_z + \Omega_k S_x,\\
\end{split}
\end{align}
where $\Delta' = \Delta_e + A_\parallel I_z + b(t)$, and $\Omega_k = \Omega_{SL} - k'\omega_f$. This reduces the Hamiltonian to a familiar form, with $\Delta'$ and $\Omega_k$ serving as effective detuning and irradiation terms on the $S$ spin. As such, $\Delta'$ will have little effect on the spin dynamics when $s_k'\Delta' \ll \Omega_k$, resulting in a decoupling of the undesired off resonance and noise effects. In addition, when $\Omega_k \approx \pm \omega_n$, with $\omega_n = \omega_n^0 + \frac{1}{2}A_\parallel$ to first order (see supplementary material (SM)), the NOVEL Hartmann-Hahn condition is met, and the $A_\perp$ part of the hyperfine interaction will result in polarization transfer between the electron and nucleus \cite{Henstra2008} (See SM). Higher $k'$ values will lead to higher quenching of the interactions, resulting in narrower resonance conditions, lower dephasing  and slower polarization transfer. We note that in the extreme case of $\Omega_{SL} = 0$ or $\omega_f \rightarrow 0$ (i.e no $\pi$ pulses) these conditions are identical to the ones used for CPMG based sensing \cite{DeVience2015} and spin-lock/NOVEL experiments, respectively. 

%%%%%%%%%%%%%%%%
% simulations
%%%%%%%%%%%%%%%%
In order to demonstrate these conditions and their effects on the spin system, we performed numerical simulations on a system composed of an NV and a single $^{13}$C nucleus, with $A_\parallel = $ 30 kHz, $A_\perp, \Delta_e =$ of 40 kHz, and an external field of 80 Gauss. The NV and nuclear polarizations, $\langle S_z \rangle_{(t)} = \langle 0_e|\rho(t)|0_e\rangle - \langle -1_e|\rho(t)|-1_e\rangle$ and $\langle I_z \rangle_{(t)} \equiv 2 Tr(I_z\rho(t))$ respectively, were calculated using the Liouville von-Neumann equation,
\begin{align}\label{LVN}
\frac{\partial\rho}{\partial t} = -i[H,\rho], 
\end{align}
where $H$ is given in Eq. \ref{Spin Hamiltonian}. When dephasing noise was considered, a bath modeled by an Ornstein-Uhlenbeck process was considered, with a correlation time $\tau_c$, and an interaction strength $\Delta_{noise}$ with the NV. This is described in more detail in the SM, based on Ref. \cite{Gillespie1996}.

\begin{figure}[!t]
	\centering   
	\includegraphics[width = 0.48\textwidth]{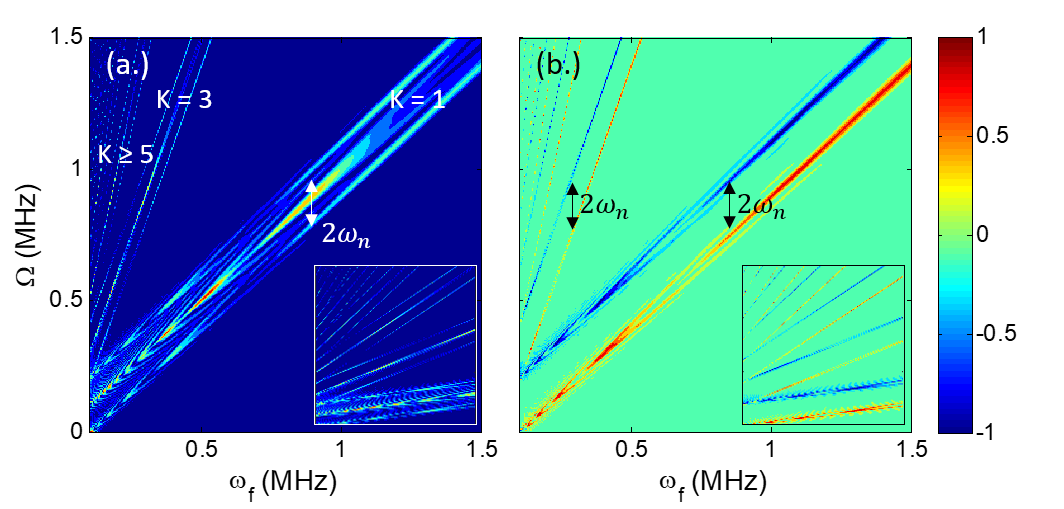}
	\caption{(a) NV polarization, $\langle S_z \rangle$, and (b) nuclear polarization, $\langle I_z \rangle$, as a function of $\omega_f$ ($\pi/\tau$) and $\Omega_{SL}$. The simulation was performed using $N = 60$, on an NV-$^{13}$C model system with $A_\parallel = $ 30 kHz, $A_\perp, \Delta_e =$ of 40 kHz, and an external field of 80 Gauss. The insets show the plot with $\omega_f$ in the range of 0.1-0.3 MHz. $\omega_n$ separation and $k$ conditions are marked schematically.}
	\label{Omega_omega_f}
\end{figure}

Fig. \ref{Omega_omega_f} depicts the resonance conditions of this sequence, with the NV and nuclear polarizations plotted as a function of the sequence periodicity $\omega_f$ and the SL power $\Omega_{SL}$. This was performed using $N = 60$ and without dephasing noise, $b(t) = 0$. Changes in $\langle S_z \rangle$ (Fig. \ref{Omega_omega_f}a) occur around $\Omega_{SL} = k\omega_f$ and $k\omega_f \pm\omega_n$ (with $\omega_n/2\pi \simeq 0.1$ MHz). The former originates from the off resonance term and $A_\parallel$, and can result in polarization inversion, but not in nuclear polarization; the latter originates from $A_\perp$, and results in NV-nuclear polarization transfer, as seen by the change in $\langle I_z \rangle$ (Fig. \ref{Omega_omega_f}b). Far from these conditions the state of the NV and nuclear polarizations remain unchanged. We note that the sign of $\langle S_z \rangle$ and $\langle I_z \rangle$ can be inverted by changing the relative phase of the final $\frac{\pi}{2}$ pulse or of the rNOVEL pulses with respect to the initial $\frac{\pi}{2}$ pulse, respectively.

\begin{figure}[!t]
	\centering   
	\includegraphics[width = 0.47\textwidth]{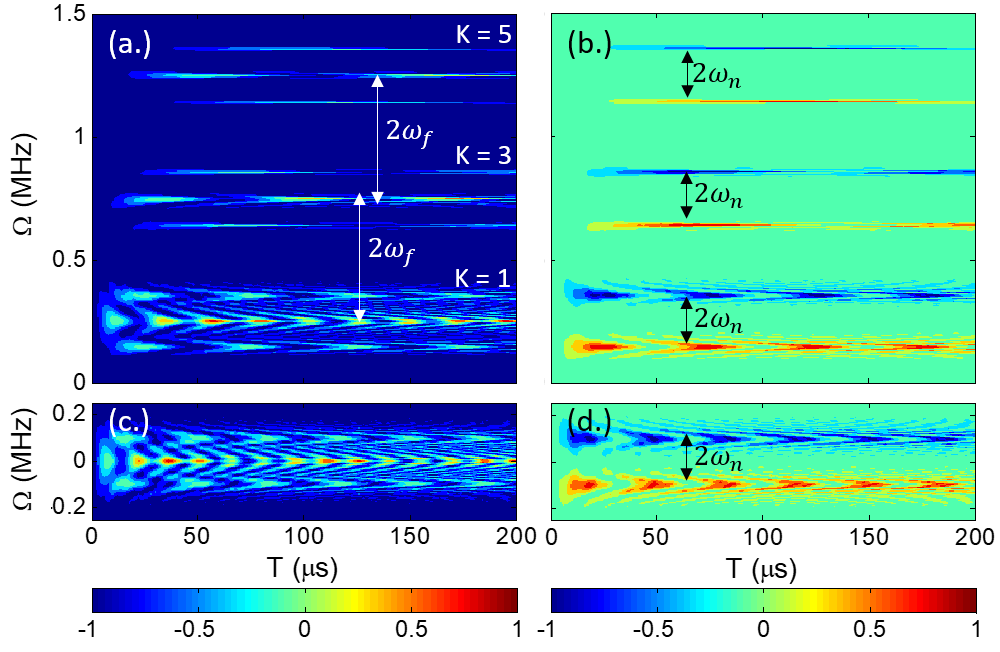}
	\caption{(a,c) $\langle S_z \rangle$ and (b,d) $\langle I_z \rangle$ as a function of $N$ and $\Omega_{SL}$, for the rNOVEL (a,b) and NOVEL (c,d) sequences. $\tau = 2\mu$s was used, resulting in a total sequence duration of $T = 2 N \mu$s. All other simulation parameters are as given in Fig. \ref{Omega_omega_f}. The $k$ = 1,3 and 5 rNOVEL conditions are marked schematically in the figure, as well as the $2\omega_f$ and $2\omega_n$ separation of the resonant conditions}
	\label{Omega_N}
\end{figure}

A more convenient form of the experiment is to sweep $\Omega_{SL}$ and $N$ with a fixed $\tau$ value (in analogy with the NOVEL sequence), resulting in discrete changes of the total experiment time $T = \tau N$. This is shown in Fig. \ref{Omega_N}(a,b), where the values of $\langle S_z \rangle$ and $\langle I_z \rangle$ are plotted, respectively, using an rNOVEL sequence with $\tau = 2 \mu s$ ($\omega_f/2\pi = 0.25$ MHz). The resonance conditions are centered around $\Omega_{SL} = 0.25, 0.75, 1.25$ MHz for the k = 1,3,5 conditions, respectively, with the nuclear polarization transfer conditions separated by $\omega_n$ from it (as was described above, and as marked in the figure). As expected, the reduction in the $s_k$ values for higher $k$ conditions (Eqs. \ref{Fourier s(t)}, \ref{omega_f rotating frame Hamiltonian}) results in slower oscillations and narrower resonance conditions. A simulation of a regular NOVEL experiment is shown in Fig. \ref{Omega_N}(c,d) for comparison. This was performed using the same parameters as the rNOVEL, but omitting the $\pi$ pulses. Here, the resonance conditions are centered around $\Omega_{SL} = 0$, with faster oscillations and broader conditions than in the rNOVEL case.

\begin{figure}[!t]
	\centering   
	\includegraphics[width = 0.47\textwidth]{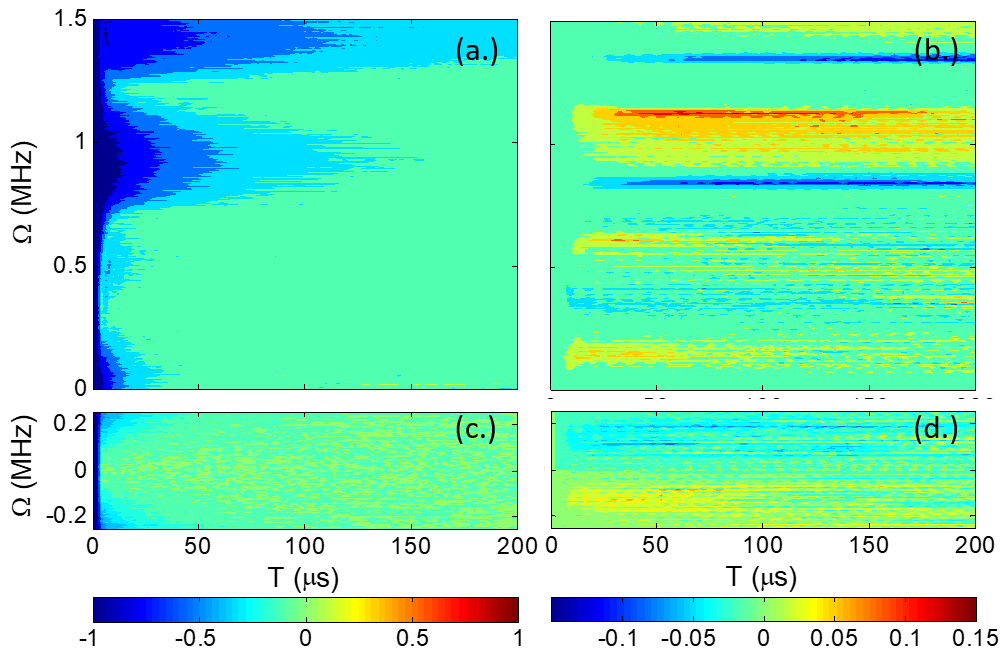}
	\caption{(a,c) $\langle S_z \rangle$ and (b,d) $\langle I_z \rangle$ as a function of $N$ and $\Omega_{SL}$ under dephasing noise, for the rNOVEL (a,b) and NOVEL (c,d) sequences.  $\tau = 2\mu$s was used, resulting in a total sequence duration of $T = 2 N \mu$s. The noise source was modeled using correlation times of $\tau_c$ = 11 and 150 $\mu$s, and $\Delta_{noise}$ of 0.5 MHz. The value of $b(t)$ was changes every 0.1 $\mu$s, as described in the SM, and the plotted $\langle S_z \rangle$ and $\langle I_z \rangle$ values were obtained by averaging over 100 occurrences. All other simulation parameters are as given in Fig. \ref{Omega_omega_f}.}
	\label{Omega_N_noise}
\end{figure}

So far we did not consider the effects of dephasing noise. We next introduce this using parameters measured for shallow NVs with a depth of about 3 nm from the diamond surface \cite{Romach2015}. This is given by fast and slow noise components, characterized by $\tau_c$ of 11 and 150 $\mu$s, and with $\Delta_{noise} =$  0.5 MHz used in both cases. The resulting NV and nuclear polarizations are plotted in Fig. \ref{Omega_N_noise}, using the same spin system as before, for the rNOVEL (a,b) and NOVEL (c,d) sequences. Note the change in $\langle I_z \rangle$ color-scale when compared to Figs. \ref{Omega_omega_f} and \ref{Omega_N}.
In the NOVEL case, the low $\Omega_{SL}$ power needed to polarize the nucleus is insufficiently strong to remove the effect of the decoherence noise. This results in a loss of NV polarization, leading to only $\sim 6\%$ maximal polarization. In the rNOVEL sequence the effect of the noise on the NV polarization is reduced with $k$, as can be seen in Fig. \ref{Omega_N_noise}(a.) when comparing the different $k$ conditions. This results in as much as $\sim13\%$ nuclear polarizaion, a factor of 2 higher than for the NOVEL case, but still lower than in the ideal scenario, in which noise was not considered (Fig. \ref{Omega_N}b). We stress that the enhanced polarization, as well as the narrow spectral response, are the significant advantages of rNOVEL compared to NOVEL, and constitute the main results of this work. Enhanced polarization could clearly benefit various applications, such as in sensing \cite{DeVience2015} and cooling \cite{london2013}, and the high spectral resolution could contribute to selective addressing \cite{Ajoy2015}.

\begin{figure}[!t]
	\centering   
	\includegraphics[width = 0.48\textwidth]{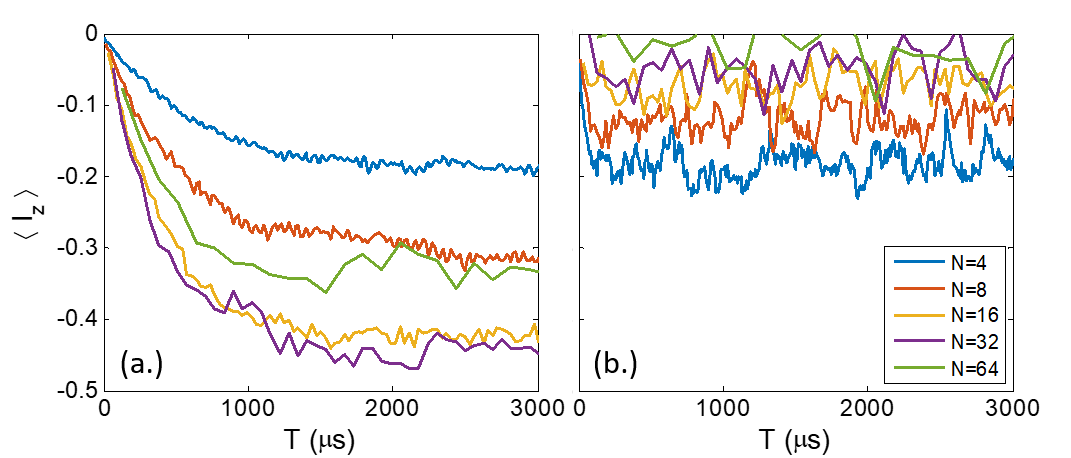}
	\caption{Nuclear polarization during rNOVEL (a) and NOVEL (b) under dephasing noise, using repeated sequence application. This is plotted as a function of the total irradiation time, for sequences composed of $N = $ 4 (blue),8 (red)),16 (yellow), 32 (purple) and 64 (green) $\pi$ pulses. $\Omega_{SL}/2\pi$ of 1.348 and 0.126 MHz were used in (a) and (b), respectively. All other parameters are the same as in Fig. \ref{Omega_N_noise}.}
	\label{cycles_noise}
\end{figure}

The limited nuclear polarization, as shown in Fig. \ref{Omega_N}, can be further increased by repeating the full sequence (Fig. \ref{sequence}) $n$ times. This was modeled as follows: first, the NOVEL/rNOVEL was applied with a given $N$. Next, ideal laser irradiation was applied, resulting in full NV decoherence and polarization (as explained in the SM). This was then followed by another NOVEL/rNOVEL sequence, and so on. The total experiment time (without the laser irradiation and $\frac{\pi}{2}$ pulses) is thus given by $T = n N \tau$. The resulting nuclear polarization,  $\langle I_z \rangle$, as a function of $T$, is plotted Fig. \ref{cycles_noise} for  rNOVEL and NOVEL using $N = $ 4, 8, 16, 32 and 64. This was performed at the rNOVEL/NOVEL polarization transfer conditions of $\Omega_{SL}/2\pi = $ 1.348 and 0.126 MHz, respectively. In the rNOVEL case the maximal nuclear polarization (and the polarization rate) increases initially with $N$, reaching its maximal value around $N = 16$ to 32. However, using higher values result in a decrease in the maximal polarization. This can be explained by a decrease in the $|\langle S_z \rangle|$ polarization (see SM), which remains mostly unchanged by the increasing nuclear polarization, and is therefore governed mainly by the dephasing term in the Hamiltonian. For the NOVEL sequence the maximal nuclear polarization decreases with $N$, again due to the loss of NV polarization (see SM). This results in lower maximal polarization when compared with the rNOVEL case. We note that the change in polarization with $N$ is a result of the conditions studied here, in which the dephasing of the NV spin is relatively significant, and thus it limits the efficiency of repeating the sequence without re-initializing the NV (especially for NOVEL). 

In real systems the NV can interact with many nuclei, including the NV's $^{14/15}$N nucleus. The strong hyperfine interaction with the latter will result in undesired off resonance terms. These can be avoided by adding $^{14/15}$N pre-polarization \cite{Pagliero2014} or pre-saturation to the initialization step. In addition, the interaction with many remote nuclei will result in faster polarization transfer from the NV to the nuclei, leading to higher NV polarization loss in a single cycle. This will limit the amount of nuclear polarization that can be transfered within a single cycle of the sequence, requiring multiple cycles for substantial nuclear polarization. In addition, this can increase the feasibility of NV based detection of nuclear polarization, even in noisy spin systems. For this goal polarization - depolarization super cycles can be used, by alternating the phase of the initial $\pi/2$ pulse phase ($y,\bar{y}$) or the rNOVEL pulses ($x,\bar{x}$) \cite{london2013}. 

While in principle the rNOVEL sequence can work at high fields, it will have disadvantages when compared with the NOVEL sequence. In particular, at high fields different re-coupling conditions can overlap, complicating the interpretation of the resulting signals. The lower limit on the field originates from the reduced separation between the noise and polarization conditions, and from the loss of nuclear polarization axis (see SM). Possible effects of the external field strength and alignment on the decoherence of the spin system must also be considered \cite{Neumann542}.    

Finally, the effects of MW imperfections were not considered here. These may be reduced using alternating phase rNOVEL sequences, in analogy with the CPMG variations such as XY-4 and XY-8 \cite{Maudsley1986, Gullion1990}, or by continuous SL phase/power modulation \cite{Cai2012, Farfurnik2017}.  

To conclude, we presented here the refocused-NOVEL sequence, which is capable of both nuclear polarization and improved noise decoupling when compared with the NOVEL sequence. This can be used even at low fields, where precise field alignment is not needed, and where the NOVEL sequence is limited due to low noise decoupling. The basic spin dynamics of the rNOVEL sequence was explained and demonstrated using numerical simulations, and we described how it can be used to tune the amount of noise decoupling, at the cost of reduced NV-nuclear polarization transfer rate. Future studies of this sequence will include experimental realization, with the goal of creating nuclear polarization at low fields. 

We thank Lucio Frydman, Dmitry Farfurnik and Hadas Inbar for fruitful discussions and
useful insights. This work has been supported in part by the Minerva ARCHES Award, the EU (ERC StG), the CIFAR-Azrieli Global Scholars program, the Ministry of Science
and Technology, Israel, and the Israel Science Foundation
(Grant No. 750/14).

\bibliography{rNOVELbib}
%%%%%%%%%%%%%%

\end{document}